\documentclass[aps, prc, nofootinbib, twocolumn]{revtex4-2}
\usepackage[colorlinks, linkcolor=red, anchorcolor=green, citecolor=blue]{hyperref}
\usepackage{amsmath}
\usepackage{amssymb}
\usepackage{graphicx}
\usepackage{color}
\usepackage{orcidlink}

\begin{document}
\title{Exact solution of Boltzmann equation in a longitudinal expanding system}
\author{Shile Chen\orcidlink{0000-0002-3874-5564}}\email{csl2023@tsinghua.edu.cn}
\affiliation{Department of Physics, Tsinghua University, Beijing 100084, China}
\author{Shuzhe Shi\orcidlink{0000-0002-3042-3093}}\email{shuzhe-shi@tsinghua.edu.cn}
\affiliation{Department of Physics, Tsinghua University, Beijing 100084, China}
\begin{abstract}
Analytical solutions to the microscopic Boltzmann equation are useful in testing the applicability and accuracy of macroscopic hydrodynamic theory.
In this work, we present exact solutions of the relativistic Boltzmann equation, based on a new family of exact solutions of the relativistic ideal hydrodynamic equations~\cite{Shi:2022iyb}. To the best of our knowledge, this is the first exact solution that allows longitudinal expansion with broken boost invariance.
\end{abstract}

\maketitle

\emph{Introduction.}
Hydrodynamics is a macroscopic, long-wavelength effective theory that describes the collective motions in many-body systems. The commonly used framework of relativistic hydrodynamic equations is derived from the microscopic Boltzmann equations by taking different approximations, such as relaxation time approximation~\cite{ANDERSON1974466}, Chapman--Enskog expansion~\cite{chapman1990mathematical, enskog}, and moment expansion based methods by Isreal--Stewart~\cite{Israel:1979wp} and Denicol \textit{et al}~\cite{Denicol:2010xn, Denicol:2012cn, Denicol:2014vaa}. 
In these derivations, the systems are assumed to be close to local equilibrium, and the distribution function is the thermal equilibrium one plus small derivations expanded in different bases. 
The hydrodynamic theory only keeps lower order moments in the Knudsen expansion, leaving the higher order ones either truncated out or extrapolated under certain assumptions.
Therefore, it is traditionally expected to be applicable in near-equilibrium systems with a small Knudsen number.

The success of applying hydrodynamic theory to describe the final state observables of collective motion in high energy nucleus-nucleus, proton-nucleus, and even proton-proton collisions (see e.g.,~\cite{Romatschke:2007mq, Song:2007ux, Schenke:2010nt, Zhao:2020pty}) has drawn significant interests in discussions of relieving the assumptions for the applicability of hydrodynamics~\cite{Heller:2011ju, Heller:2013fn,  Heller:2015dha, Kurkela:2015qoa, Blaizot:2017lht, Romatschke:2017vte, Spalinski:2017mel, Strickland:2017kux, Romatschke:2017acs, Behtash:2017wqg, Blaizot:2017ucy, Romatschke:2017ejr, Kurkela:2018wud, Mazeliauskas:2018yef,  Behtash:2019txb, Heinz:2019dbd, Blaizot:2019scw, Blaizot:2020gql, Blaizot:2021cdv}. A crucial step is to compare the hydro results to those of the Boltzmann equation, which describes the evolution of the microscopic particle distribution.

Taking the general form in a curvilinear coordinates, the Boltzmann equation for on-shell distribution function reads, (see e.g., ~\cite{Cercignani2002, Hohenegger:2008zk, Lee:2012sd})
\begin{align}
\begin{split}
&
    \Big( p^\mu\partial_\mu +  {\Gamma^\rho}_{\mu\nu} p^\mu  p_\rho \frac{\partial}{\partial  p_\nu}\Big) f( x^\alpha,  p_\beta) = \mathcal{C}[f]\,, \\
&
    \mathcal{C}[f]=
    \frac{ p_\mu  u^\mu( x)}{\tau_{r}( x)}\Big(f( x^\alpha,  p_\beta)-f_{eq}( x^\alpha,  p_\beta)\Big)\,,
\end{split}
\label{eq:boltzmann}
\end{align}
where we have taken the relaxation time approximation(RTA)~\cite{ANDERSON1974466} for the collisions kernel. The Christoffel symbol is given by
${\Gamma^\rho}_{\mu\nu} = \frac{g^{\rho\gamma}}{2}\Big(\frac{\partial  g_{\gamma\mu}}{\partial  x^{\nu}}+\frac{\partial  g_{\gamma\nu}}{\partial  x^{\mu}}-\frac{\partial  g_{\mu\nu}}{\partial x^{\gamma}}\Big)$, with $g^{\mu\nu}$ being the metric.
Taking the conformal limit that $m=0$, the relaxation time reads $\tau_{r}(\hat x) = 5\bar{\eta}/\hat T(\hat x)$, and $\bar\eta$ is the shear viscosity to entropy ratio~\cite{Denicol:2010xn, Denicol:2012cn, Florkowski:2013lya, Florkowski:2013lza}. The local equilibrium distribution function is assumed to be of Boltzmann form, $f_{eq} = \exp(-\hat p_\mu \hat u^\mu/\hat T)$. For the distribution functions in Eq.~\eqref{eq:boltzmann}, all momentum arguments are covariant and coordinate arguments are contravariant.
$\hat T(\hat x)$ and $\hat u^\mu(\hat x)$ are respectively the space-time dependent temperature and velocity. The space-time profiles of $\hat T$ and $\hat u^\mu$ are not arbitrary --- they are constrained by continuity relations such as hydrodynamics.

In a given coordinate system, if one makes certain assumptions such that the Boltzmann equation can be expressed in a form without the derivatives with respect to the spatial coordinates and momenta, there exists a formal solution~\cite{Baym:1984np, Florkowski:2013lya, Florkowski:2013lza}. With explicit form which will be shown later in the main text, the formal solution depends only on the temporal coordinate and is homogeneous with respect to the spatial one. The non-trivial spatial expansion can be introduced if one takes a coordinate of a co-moving frame of a known solution to the hydrodynamic equations~\cite{Denicol:2014xca, Denicol:2014tha, Strickland:2018ayk, Chattopadhyay:2021ive}. 
A family of exact solutions for ideal fluids is found, recently, in~\cite{Shi:2022iyb}, which is homogeneous in the transverse plane and allows expansion --- either symmetric or asymmetric --- in the longitudinal direction which breaks the boost invariance. In this work, we find the co-moving frame of the new solution and construct the formal solution accordingly. We then compute the hydrodynamics quantities and analyze how they relax to the hydro limit.

\vspace{3mm}
\emph{Co-moving frame of the longitudinal expanding flow.}
Taking the general form in a curvilinear coordinates, the hydrodynamic equation reads
\begin{align}
\mathcal D_\mu T^{\mu\nu} \equiv \partial_\mu T^{\mu\nu} +\Gamma^\mu_{\ \rho\mu}T^{\rho\nu}+\Gamma^\nu_{\ \rho\mu} T^{\rho\mu} = 0,
\end{align}
where $T^{\mu\nu}$ is the energy-momentum tensor, $\mathcal D_\mu$ is the covariant derivative. In Ref.~\cite{Shi:2022iyb}, a family of exact solutions for longitudinally expanding ideal fluids was found,
\begin{align}
\begin{split}
\frac{T_\mathrm{ideal}}{T_i} =&\; 
	\bigg( 
	\frac{t_0}{\tau_i} + 
	\frac{a\,\tau e^{\eta}}{\tau_i} \bigg)^{\frac{1-c_s^2}{4}\frac{1}{a^2}-\frac{1+c_s^2}{4}} 
\\&\;\times
	\bigg(
	\frac{t_0}{\tau_i} + 
	\frac{\tau e^{-\eta}}{a\,\tau_i}
	\bigg)^{\frac{1-c_s^2}{4} a^2-\frac{1+c_s^2}{4}}
\,,\\
u^\tau =&\; \frac{1}{2} \Bigg(
	\sqrt{ \frac{t_0 e^{-\eta} + \tau\, a}{t_0 e^{\eta} + \tau/a} }
+	\sqrt{ \frac{t_0 e^{\eta} + \tau/a}{t_0 e^{- \eta} + \tau\, a} }
 \Bigg)\,,\\
u^\eta =& \frac{1}{2\tau} \Bigg(
	\sqrt{ \frac{t_0 e^{-\eta} + \tau\, a}{t_0 e^{\eta} + \tau/a} }
-	\sqrt{ \frac{t_0 e^{\eta} + \tau/a}{t_0 e^{- \eta} + \tau\, a} }
 \Bigg).
\end{split} \label{eq:hydro:solution.1}
\end{align}
In the solution, $\tau$ and $\eta$ are the proper-time and rapidity in Milne coordinates, respectively. $c_s$ is the speed of sound, $a$ is a dimensionless parameter characterizing the asymmetry between forward and backward rapidity range, $T_i$ and $\tau_i$ are positive constants that respectively scale the temperature and time. The non-negative time constant $t_0$ serves as a translation of the Minkowski time, and it corresponds to the time needed for the colliding nuclear pancakes to pass through each other in relativistic heavy-ion collisions~\cite{Shi:2022iyb}. Focusing on the central rapidity slice for a reflectional symmetric system ($a=0$), the temperature decreases as $T(\tau) = T_i \,\big(\frac{\tau_i}{\tau+t_0}\big)^{\frac{1}{3}}$, which decays slower than the Bjorken flow at small $\tau$ and approaches the Bjorken limit when $\tau \gg t_0$.

In this work, we introduce a new coordinate system which can be transformed from the Milne coordinates 
\begin{align}
\begin{split}
\hat{x}^0 =\;& 
    \frac{2a\,\tau_i}{1+a^2}\Big(
    \big(\frac{t_0 + a\,\tau e^{\eta}}{\tau_i}\big)^{\frac{1}{a}}
    \big(\frac{{t}_0}{\tau_i} + \frac{\tau e^{-\eta}}{a\,\tau_i}\big)^a 
    \Big)^{\frac{1+a^2}{4a}} \,,\\
\hat{x}^1 =\;& 
    \frac{1+a^2}{4a}\ln\Big(
    \big(\frac{t_0 + a\,\tau e^{\eta}}{\tau_i}\big)^{\frac{1}{a}}
    \big/
    \big(\frac{{t}_0}{\tau_i} + \frac{\tau e^{-\eta}}{a\,\tau_i}\big)^a 
    \Big) \,,\\
\hat{x}^x =\;& x,\qquad\qquad \hat{x}^y=y.
\end{split}
\label{eq:coordinate}
\end{align}
The corresponding metric is 
\begin{align}
\hat g^{\mu\nu} = \mathrm{diag}\Big[
    e^{2\frac{1-a^2}{1+a^2} \hat{x}^1},
    - (\hat{x}^0)^{-2}
    e^{2\frac{1-a^2}{1+a^2} \hat{x}^1},
    -1,-1\Big]\,,
\end{align}
and the non-vanishing components of the Christoffel symbol are
\begin{align}
\begin{split}
&   
    \hat{\Gamma}^{0}_{\;\,11} = \hat{x}^0\,,\quad
    \hat{\Gamma}^{1}_{\;\,10} = \hat{\Gamma}^{1}_{\;\,01} = \frac{1}{\hat{x}^0}\,,
\\
&    \hat{\Gamma}^{0}_{\;\,01} = \hat{\Gamma}^{0}_{\;\,10} 
=    \hat{\Gamma}^{1}_{\;\,11} = (\hat{x}^0)^2\hat{\Gamma}^{1}_{\;\,00}
=    \frac{a^2-1}{a^2+1}\,.
\end{split}
\end{align}
From now on, we take the ``hat'' ($\hat{\cdot}$) notation to denote quantities under the new coordinate system~\eqref{eq:coordinate}. Under the co-moving frame, the solution~\eqref{eq:hydro:solution.1} becomes ``static'' that all spatial components vanish, and the space-time profile of the solution reads
\begin{align}
\begin{split}
&\hat T_\mathrm{ideal} = 
     \sqrt{\hat{g}^{00}}\,T_i\,\Big(\frac{\tau_i}{\hat{x}^0}\Big)^{c_s^2}\,,\\
&\hat u^0 =\sqrt{\hat{g}^{00}}\,,\quad
\hat u^1 = \hat u^x = \hat u^y = 0\,.
\end{split}\label{eq:hydro:solution.2}
\end{align}
Therefore, \eqref{eq:coordinate} is the the co-moving frame of the asymmetric expanding flow~\eqref{eq:hydro:solution.1}. We have redefined $T_i$ to absorb a space-time independent factor.

Taking $t_0 = 0$ and $a = 1$, Eq.~\eqref{eq:coordinate} returns to the Milne coordinate and Eq.~\eqref{eq:hydro:solution.2} returns to the Bjorken--Hwa solution~\cite{Hwa:1974gn, Bjorken:1982qr}. 
With general values for $t_0$ and $a$, one may still connect the new solution with the Bjorken--Hwa solution by re-scaling $\hat{u}^\mu$ and $\hat{T}$ by $(\hat{g}^{00})^{-1/2}$ and replacing $\hat{x}^0$ by $\tau$. 
$\hat{x}^0$ and $\hat{x}^1$, thus, are respectively referred to as the hat-proper-time and hat-rapidity in this paper.

\vspace{3mm}
\emph{The Boltzmann equation in the co-moving frame.}
Noting the simplicity of the solution~\eqref{eq:hydro:solution.2} in the co-moving frame~\eqref{eq:coordinate} and its similarity to the Bjorken--Hwa solution, we take the co-moving frame and solve the Boltzmann equations.
Following the property of the hydro solution, we focus on the systems that are homogeneous in the transverse plane, and the Boltzmann equation~\eqref{eq:boltzmann} in the co-moving frame becomes
\begin{align}
\Big(\hat{g}^{00}\hat{p}_0\frac{\partial}{\partial{{\hat x}^0}}+ \hat{g}^{11}\hat{p}_1\frac{\partial}{\partial{{\hat x}^1}} +\frac{a^2-1}{a^2+1} (\hat p_x^2+\hat p_y^2) \frac{\partial}{\partial{{\hat p}_1}}\Big)f(\hat x^\alpha,\hat p_\beta) \nonumber\\
= -\frac{\hat{p}_0 \hat{u}^0 \hat{T}(\hat{x}^0)}{5\bar{\eta}}\Big(f(x^\alpha,\hat p_\beta)-f_{eq}(x^\alpha,\hat p_\beta)\Big).
\label{eq:boltzmann.2}
\end{align}
Noting that the solution~\eqref{eq:hydro:solution.1} requires a simple relation between the pressure and the energy density, $P = c_s^2\varepsilon$, which corresponds to the conformal limit in a kinetic theory, we therefore focus on massless particles in this work.
The non-vanishing prefactor of the derivative with respect to ${\hat p}_1$ posts challenges in getting exact solution of the Boltzmann equation: the formal solution~\cite{Baym:1984np, Florkowski:2013lya, Florkowski:2013lza} can no longer be applied in such cases. 
One may try to simplify the coordinate and momentum dependence of the distribution function by analyzing the symmetry properties of the hydro solution~\eqref{eq:hydro:solution.2} and the Boltzmann equation~\eqref{eq:boltzmann.2}. We note that~\eqref{eq:boltzmann.2} is invariant if one performs a pseudo boost that $\hat x^0 \to \hat x^0$ and $\hat x^1 \to \hat x^1 + \eta_b$, $\hat x^i \to \hat x^i\,e^{-\frac{1-a^2}{1+a^2}\eta_b}$ for $i \in \{x, y\}$, 
which leads to $\hat g^{\mu\nu} \to e^{2\frac{1-a^2}{1+a^2}\eta_b} \hat g^{\mu\nu}$, and correspondingly, $\hat p_T \to \hat p_T\,e^{\frac{1-a^2}{1+a^2}\eta_b}$, $\hat T \to \hat T\,e^{\frac{1-a^2}{1+a^2}\eta_b}$, $\hat u^0 \to \hat u^0\,e^{\frac{1-a^2}{1+a^2}\eta_b}$. Noting that such a transformation scales the transverse coordinates differently than the temporal and longitudinal ones, it simplify~\eqref{eq:boltzmann.2} under two special situations. One is to eliminate the $\hat{x}^x$ and $\hat{x}^y$ coordinates; the other is to let $a=1$ so that the transverse scaling factor ($e^{\frac{1-a^2}{1+a^2}\eta_b}$) is unity. Both simplifications correspond to special solutions of~\eqref{eq:boltzmann.2} and will be discussed in what follows sequentially.

\vspace{3mm}
\emph{Solution in 1+1 D.}
First, we consider a coordinate system with only $\hat{x}^0$ and $\hat{x}^1$, i.e., a $1+1$ dimensional system.
This is equivalent to considering $f \propto \delta(\hat p_x)\delta(\hat p_y)$ so that momentum derivative in~\eqref{eq:boltzmann.2} is eliminated. 
Under such a constraint, the on-shell condition becomes $\hat p_0 = {|\hat p_1|}/{\hat x^0} $, 
and the speed of sound for massless free particle is $c_s=1$. We find that both the homogeneous equilibrium distribution,  and the equivalent relaxation time are independent of $\hat x^1$. 
This further allows us to assume the solution to be independent of the hat-rapidity, and Eq.~\eqref{eq:boltzmann.2} becomes
\begin{align}
    \partial_{\hat x^0} f(\hat x^0,\hat p_1) 
= 
    -\frac{\hat T({\hat x}^0)}{5\bar{\eta}}
    \Big(f(\hat x^0,\hat p_1) - \exp(-\frac{\hat p_0}{\hat T(\hat x^0)})\Big).
\label{eq:boltzmann.3}
\end{align}
In 1+1D, $\bar{\eta}$ has no physics meaning, but it is a dimensionless parameter that scales the relaxation time.
With arbitrary initial distribution function given at hat-proper-time $\hat x^0_i$, $f_i(\hat x^0_i, \hat p_1)$, based on Landau matching, we find the equation to the effective temperature of 1+1D system
\begin{align}
\begin{split}
\hat T_\mathrm{eff}^2({\hat x}^0)=\;&\frac{({\hat x}^0)^2}{({\hat x}^0_i)^2}e^{-\frac{1}{5\bar \eta}\int_{{\hat x}_i^0}^{{\hat x}^0}\hat T_\mathrm{eff}(x')dx'}\hat T_0^2 
\\+\;& \frac{1}{5\bar\eta}\int_{{\hat x}_i^0}^{{\hat x}^0}dx' \frac{({\hat x}^0)^2}{({\hat x}')^2} e^{-\frac{1}{5\bar \eta}\int_{{\hat x}'}^{{\hat x}^0}\hat T_\mathrm{eff}(x'')dx''}\hat T_\mathrm{eff}^3({\hat x}')\,.
\end{split}
\end{align}
When $\bar \eta\to 0$, effective temperature returns to ideal hydro solution $\hat T({\hat x}^0) = T_i\, \tau_i/{\hat x}^0$.

It shall be worth noting that while $f(\hat x^0, \hat p_1)$ does not explicitly depend on hat-rapidity, the corresponding energy-momentum stress tensor still non-trivially depends on $\hat x^1$ via the Jacobian of the momentum integral (i.e., $\sqrt{\hat{g}^{00}}$). In other words, the $\hat x^1$-dependence of hydrodynamic quantities can be factored out as scaling constants.

\vspace{3mm}
\emph{Symmetric solution in 3+1D.}
Compared to the aforementioned $1+1$ D solution, the longitudinal distribution of relativistic heavy-ion collisions particle production is better described by solutions in the 3+1 dimensional coordinate~\cite{Shi:2022iyb}, even if the transverse profile are assumed to be homogeneous. It is, thus, important to find solutions of the Boltzmann equation in $3+1$ D coordinates for realistic studies.
When taking into account the transverse degrees of freedom, a solution of the Boltzmann equation~\eqref{eq:boltzmann.2} can be found only when $a=1$, which describes systems that are symmetric when reflecting the longitudinal direction.
When $a=1$, the hydro solution~\eqref{eq:hydro:solution.2} is equivalent to the Bjorken--Hwa flow if one performs a mapping between the hat and Milne coordinates, $\hat x^0 \leftrightarrow \tau$ and $\hat x^1 \leftrightarrow \eta$. Such a mapping is essentially a Minkowski-time translation~\cite{Shi:2022iyb}, but it corresponds to non-trivial physics consequence as it gives the correct plateau structure in the rapidity distribution of final state particles which is observed in experiments. The coordinate mapping naturally leads to a corresponding relation between solution to the Boltzmann equation and ~\cite{Chattopadhyay:2021ive}.
The Boltzmann equation takes a simple form,
\begin{align}
    \partial_{\hat x^0} f(\hat x^0,\hat p_\mu)
=
    -\frac{\hat T(\hat x^0)}{5\bar{\eta}}
    \Big(f(\hat x^0,\hat p_\mu) - e^{-\frac{\hat p_0(\hat x^0)}{\hat T(\hat x^0)}}\Big),
\end{align}
and the formal solution~\cite{Baym:1984np, Florkowski:2013lya, Florkowski:2013lza} is applicable,
\begin{align}
\label{distributionf}
\begin{split}
    f(\hat x^0, \hat p_1, \hat p_T) 
=\;&  
    D(\hat x^0, \hat x^0_i)f_0(\hat x^0_i, \hat p_1, \hat p_T)\\
+\;&
    \frac{1}{5\bar{\eta}} \int_{\hat x^0_i}^{\hat x^0} d{\hat x}' D(\hat x^0, {\hat x}' ) \hat{T}({\hat x}')e^{-\frac{\hat p_0(\hat x')}{\hat T(\hat x')}}\,.
\end{split}
\end{align}
Here, $f_0(\hat x^0_i, \hat p_1, \hat p_T)$ is the initial distribution at hat-proper-time $\hat x^0_i$, $D(\hat x^0, \hat x^0_i) \equiv
\exp\Big[-\frac{1}{5\bar{\eta}}\int_{\hat x^0_i}^{\hat x^0} \hat{T}({\hat x}') d{\hat x}'\Big]$, and the energy is given by the on-shell condition $\hat p_0(\hat x^0) = \sqrt{(\frac{\hat p_1}{\hat x^0})^2 + \hat p_T^2}$. The temperature is fixed by ensuring energy conservation, $0 = \int d\hat{P} (\hat{p}^0)^2 (f(\hat x^0,\hat p_\mu) - e^{-\frac{\hat p_0}{\hat T(\hat x^0)}})$, where $\int d\hat{P} \equiv \int  \frac{d^4 \hat p}{(2\pi)^3\sqrt{-\hat g}}2\delta((\hat p_0)^2-(\frac{\hat p_1}{\hat x^0})^2 - \hat p_T^2)$ with $\sqrt{-\hat g} = \sqrt{-\det \hat g_{\mu\nu}} = \hat x^0$ and $d^4\hat p = d\hat p_0d\hat p_1d\hat p_xd\hat p_y$ . There is no simple explicit form for the integral of the solution~\eqref{distributionf}. 
Nevertheless, we may solve the integral numerically, construct the stress tensor out of the distribution function, extract the macroscopic hydrodynamic quantities, and analyze their ``time'' evolution.


\vspace{3mm}
\emph{Hydrodynamic quantities.}
Given the distribution function, the energy momentum tensor reads
\begin{align}
\label{Tmunuf}
    \hat T^{\mu\nu} 
= 
    \int d\hat P\, \hat p^\mu\hat p^\nu f(\hat x,\hat p).
\end{align}
In the conformal limit, the bulk viscosity vanishes, and one can always decompose stress tensor as
\begin{align}
    \hat T^{\mu\nu} 
= 
    \hat \varepsilon(\hat x) \hat u^\mu\hat u^\nu+\hat  \Delta^{\mu\nu}\hat{\mathcal P}(\hat x)
    + \hat \pi^{\mu\nu}(\hat x)\,,
\end{align}
where $\hat \varepsilon(\hat x)$ is the energy density, $\hat {\mathcal P}(\hat x)$ the thermodynamic pressure, $\hat \pi^{\mu\nu}(\hat x)$ the shear viscous stress tensor, and $\hat\Delta^{\mu\nu}\equiv \hat g^{\mu\nu} - \hat u^\mu\hat u^\nu$ the ``space'' projection operator. $\hat \pi^{\mu\nu}$ is symmetric, traceless ($\hat{g}_{\mu\nu} \hat \pi^{\mu\nu} = 0$), and orthogonal to the fluid velocity ($\hat{u}_{\mu} \hat \pi^{\mu\nu} = \hat{u}_{\nu} \hat \pi^{\mu\nu} = 0$). In 3+1 D, a conformal system has $c_s = 1/\sqrt{3}$, and the energy density, thermodynamic pressure, and entropy only depend on temperature
\begin{align}
    \hat\varepsilon = \frac{3\, \hat T^4}{\pi^2}\,,\quad
    \hat P = \frac{\hat T^4}{\pi^2}\,,\quad
    \hat s = \frac{\hat\varepsilon + \hat P}{\hat T} = \frac{4\,\hat T^3}{\pi^2}\,.
\label{eq:eos}
\end{align}

\begin{figure}[!thb]
    \centering
    \includegraphics[width=0.45\textwidth]{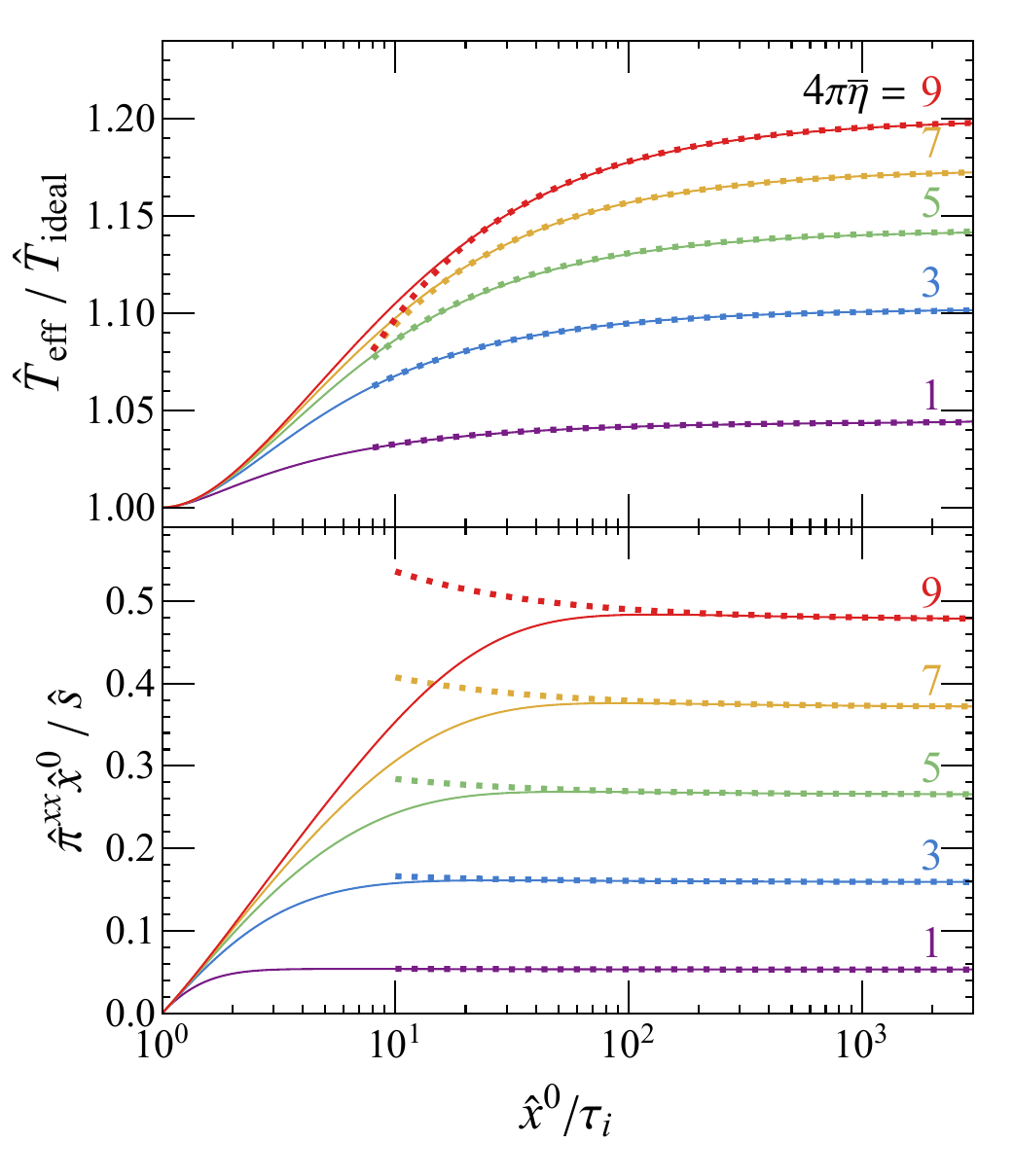}
    \caption{Scaled temperature (upper) and scaled temperature shear viscous stress tensor (lower) as functions of $\hat x^0/{\hat x}_i^0$ obtained by different shear viscosity to entropy ratios ($\bar\eta$).
    Curves are respectively for $\bar \eta = 1/4\pi$(purple), $3/4\pi$(blue), $5/4\pi$(green), $7/4\pi$(yellow), and $9/4\pi$(red).
    Dashed curves correspond to the long-time asymptotic solutions~\protect{\eqref{eq:longtime}}.
    \label{fig1}}
\end{figure}

\begin{figure*}[!hbtp]
    \centering
    \includegraphics[width=0.45\textwidth]{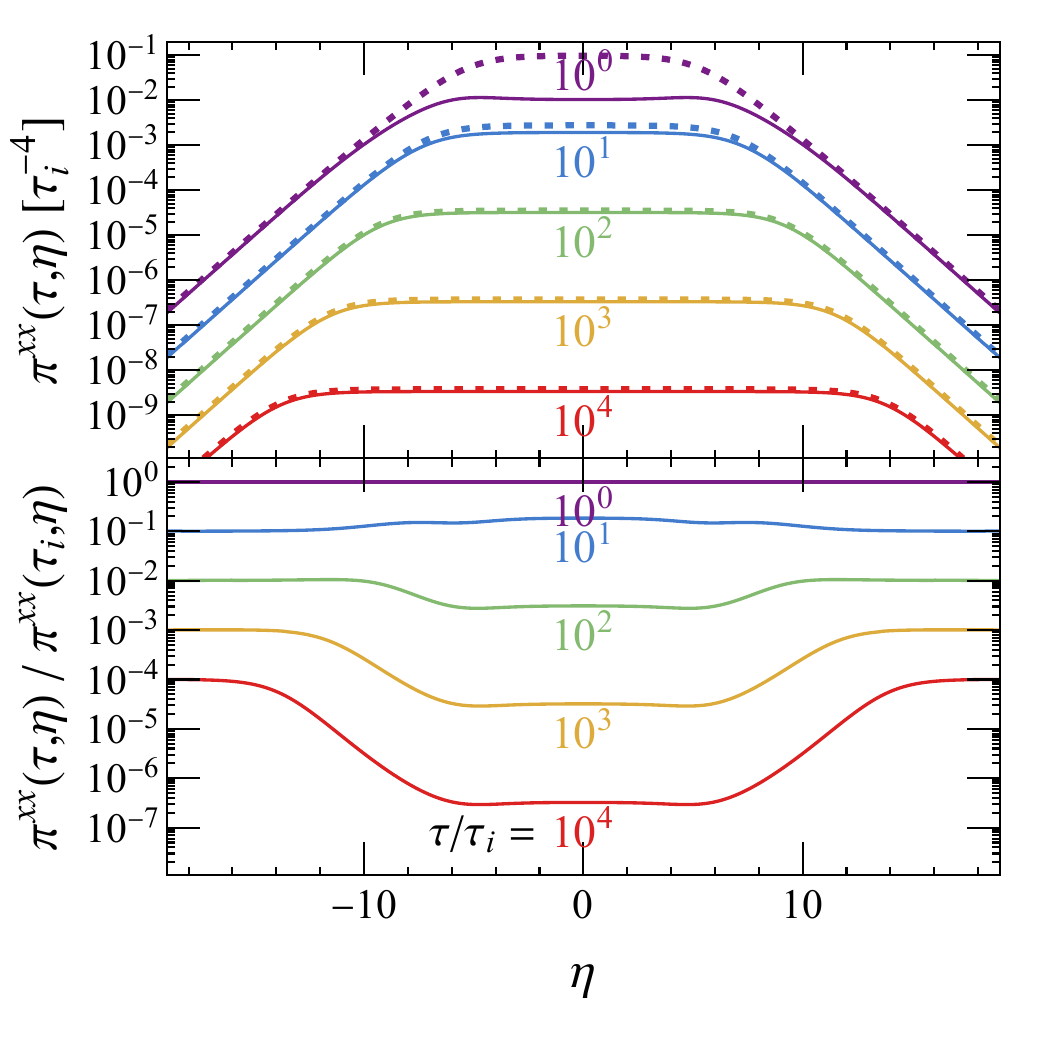}
    \includegraphics[width=0.45\textwidth]{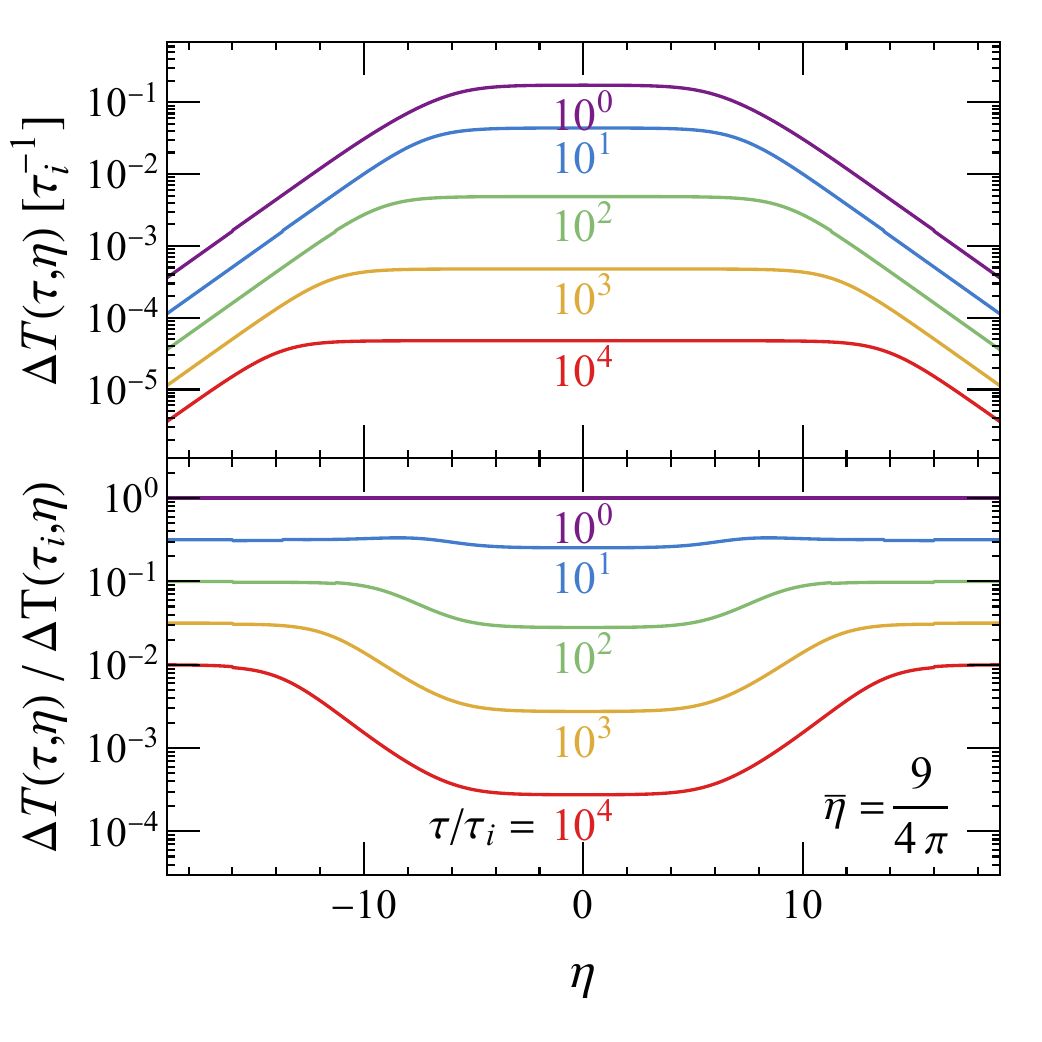}
    \caption{Rapidity dependence of shear viscous tensor (left) and the difference between the scaled ideal solution the actual temperature (right) at proper time $\tau/\tau_i = 1$(purple), $10$(blue), $10^2$(green), $10^3$(yellow), and $10^4$(red). Upper panels are the absolute values whereas lower panels are the ratios between the value at any proper time to those at initial time $\tau=\tau_i$. In the upper left panel, dashed curves are added indicating the Navier--Stokes solution that ${\pi}^{xx} = \frac{2\bar\eta}{3}\frac{{s}}{\hat x^0}$.
    \label{fig2}}
\end{figure*}
Placing $\int d\hat{P}\, \hat p^\mu \hat p^\nu$ on both side of Eq.~\eqref{distributionf}, we obtain the integration form of the stress tensor,
\begin{align}
\begin{split}
 \;&    \hat{T}^{\mu\nu}(\hat x^0)
=
    D(\hat x^0, \hat x^0_i) 
    \mathcal{H}^{\mu\nu}(\frac{\hat x^0_i}{\hat x^0})
    \int \frac{p^3 dp}{2\pi^2} f_0(\hat x^0_i,\hat p_{1},\hat p_{T})
\\+\;&
    \frac{1}{5\bar{\eta}} \int_{\hat x^0_i}^{\hat x^0} d{\hat x}' D(\hat x^0, {\hat x}' ) \hat{T}({\hat x}') 
    \mathcal{H}^{\mu\nu}(\frac{\hat x'}{\hat x^0})
    \int \frac{p^3 dp}{2\pi^2} f_{eq}(\hat x',\hat p_0(\hat x')),
\end{split}
\label{eq:Tmunu}
\end{align}
with non-vanishing elements of tensor $\mathcal H^{\mu\nu}$ given by
\begin{align}
\begin{split}
\mathcal H^{00}(\xi) \equiv\;&
    \frac{\xi}{4\pi}
    \int \frac{\sin\theta d\theta d\varphi}{(\sin^2\theta + \xi^2\cos^2\theta)^{-\frac{1}{2}}}=\mathcal R(\xi)\,,\\
\mathcal H^{11}(\xi) \equiv\;& 
    \frac{\xi^3}{4\pi}\frac{1}{(\hat x^0)^2}
    \int  \frac{\cos^2\theta\sin\theta d\theta d\varphi}{(\sin^2\theta + \xi^2\cos^2\theta)^{\frac{1}{2}}}=\frac{\mathcal R^L(\xi)}{(\hat x^0)^2}\,,\\
\mathcal H^{xx}(\xi) \equiv\;&
    \frac{\xi}{4\pi}
    \int  \frac{\sin^2\theta\cos^2\varphi \sin\theta d\theta d\varphi}{(\sin^2\theta + \xi^2\cos^2\theta)^{\frac{1}{2}}}=\mathcal R^T(\xi)\,,
\end{split}
\end{align}
where $\mathcal R(\xi) = \frac{\xi}{2}(\xi+\frac{\arccos{\xi}
}{\sqrt{1-\xi^2}})$, $\mathcal R^L(\xi) = \frac{\xi^2(\mathcal R(\xi)-\xi^2)}{1-\xi^2}$, $\mathcal R^T(\xi) = \frac{(1-2\xi^2)\mathcal R(\xi)+\xi^4}{2(1-\xi^2)}$, and $\mathcal H^{yy}(\xi) = \mathcal H^{xx}(\xi)$~\cite{Alqahtani:2017mhy}.
Noting that $\hat T^{\mu\nu}$ is always diagonal, the fluid is still ``at rest'' in the co-moving frame, i.e., $\hat{u}^\mu$ in~\eqref{eq:hydro:solution.2} remains the correct velocity decomposition of~\eqref{eq:Tmunu}.
For a conformal system that is homogeneous in the transverse plane, the stress tensor is traceless ($\hat g_{\mu\nu}\hat T^{\mu\nu} = 0$) and symmetric when exchanging the transverse variables ($\hat T^{xx}=\hat T^{yy}$).
There remains two independent components in the stress tensor --- the effective temperature determined by $\hat T^{00} = \frac{3\,\hat{T}_\mathrm{eff}^4}{\pi^2}$, and the shear viscous tensor $\hat\pi^{xx} \equiv \hat{T}^{xx} - \frac{\hat T^{00}}{3}$.

We start the evolution from an equilibrium distribution with temperature denoted by $\hat T_0$, and
the effective temperature and shear viscous stress tensor read
\begin{align}
\begin{split}
    \hat T_\mathrm{eff}^4(\hat x^0)
=\;& 
    D(\hat x^0, \hat x^0_i)\tilde{\mathcal{H}}_0 (\frac{\hat x^0_i}{\hat x^0}) \hat T^4_0
\\+\;&
    \frac{1}{5\bar{\eta}} \int_{\hat x^0_i}^{\hat x^0} d{\hat x}' D(\hat x^0, {\hat x}' ) \tilde{\mathcal{H}}_0 (\frac{\hat x'}{\hat x^0}) \hat T_\mathrm{eff}^5(\hat x')\,,
\\
    \hat \pi^{xx}(\hat x^0)
=\;& 
    D(\hat x^0, \hat x^0_i)
    \tilde{\mathcal{H}}_x(\frac{\hat x^0_i}{\hat x^0}) 
    \hat T^4_0
\\+\;&
    \frac{1}{5\bar{\eta}} \int_{\hat x^0_i}^{\hat x^0} d{\hat x}' D(\hat x^0, {\hat x}') \tilde{\mathcal{H}}_x (\frac{\hat x'}{\hat x^0}) \hat T_\mathrm{eff}^5(\hat x')\,,
\end{split}\label{eq:T_and_pi}
\end{align}
where $\tilde{\mathcal{H}}_0 \equiv \mathcal{H}^{00}$ and $\tilde{\mathcal{H}}_x \equiv \frac{3}{\pi^2}(\mathcal{H}^{xx}-\frac{\mathcal{H}^{00}}{3})$. We therefore look into the dimensionless quantities, scaled temperature $\hat T_\mathrm{eff} / \hat T_\mathrm{ideal}$ and scaled shear viscous stress tensor ${\hat x^0 \hat \pi^{xx}}/{\hat s}$, whose ``evolution" with hat-proper-time are shown in Fig.~\ref{fig1} for various $\bar\eta$. 
Matching condition of the initial state requires that parameters in $\hat T_\mathrm{ideal}$ shall be taken as $T_i = \hat T_0$ and $\tau_i = \hat x^0_i$.
In making the plot, we have set $\hat T_0 = 1/\tau_i$.

In Fig.~\ref{fig1}, dashed curves are also shown to indicate the long hat-proper-time asymptotic analytical results obtained from a perturbative analysis of~\eqref{eq:T_and_pi},
\begin{align}
\begin{split}
    \frac{\hat{T}_\mathrm{eff}}{\hat{T}_\mathrm{ideal}}
=\,&
    e_{\bar\eta} 
    - \frac{2\,\bar{\eta}}{3}
        \Big(\frac{\tau_i}{\hat x^0}\Big)^{\frac{2}{3}}
    + \mathcal{O}\big((\hat x^0)^{-\frac{4}{3}}\big)
    \,,\\
    \frac{\hat{\pi}^{xx} \hat x^0}{\hat s}
=\,&
    \frac{2\bar{\eta}}{3} 
    + \frac{40\,\bar{\eta}^2}{63\, e_{\bar\eta}\,T_i\,\tau_i} \Big(\frac{\tau_i}{\hat x^0}\Big)^{\frac{2}{3}}
    + \mathcal{O}\big((\hat x^0)^{-\frac{4}{3}}\big)\,.
\end{split}\label{eq:longtime}
\end{align}
Details are shown in Appendix.
Starting from an initial condition which matches the ideal temperature, the scaled temperature first deviates from unity driven by the viscous effect, then it approaches the ideal limit with the effective temperature being $e_{\bar\eta} T_i$ at large enough $\hat x^0$. $e_{\bar\eta}$ is a viscosity-dependent constant that shall be obtained numerically, and it is greater than unity due to entropy production. The scaled shear-viscous stress tensor starts from zero and approaches the Navier--Stokes limit. Meanwhile, the longitudinal to transverse pressure ratio is given by $\frac{P_L}{P_T} = \frac{\hat P-2\hat \pi^{xx}}{\hat P+\hat \pi^{xx}} = 1 - \frac{3\,\bar\eta}{2\,e_{\bar{\eta}} T_i \tau_i} \big(\frac{\tau_i}{\hat x^0}\big)^{\frac{2}{3}} + \mathcal{O}\big((\hat x^0)^{-\frac{4}{3}}\big)$.

Finally, we are ready to study the proper-time and rapidity dependence of temperature and shear-viscosity by transforming back to the Milne coordinates, $T(\tau,\eta) = \hat{T}(\hat{x}^0(\tau,\eta))$ and $\pi^{xx}(\tau,\eta) = \hat{\pi}^{xx}(\hat{x}^0(\tau,\eta))$. In Fig.~\ref{fig2}, we show the rapidity dependence of shear viscous tensor and the difference between the temperature and the corresponding long-time ideal hydro limit, $\Delta T \equiv |\hat T - e_{\bar\eta} \hat T_\mathrm{ideal}|$. We have set the overlap time $t_0 = 0.01~\mathrm{fm}/c$ as in~\cite{Shi:2022iyb}. In a particular $\eta$ slices, the solution is approaching to a ideal hydro limit when proper time increase --- values of both $\pi^{xx}$ and $\Delta T$ are getting smaller in $\tau$. $\pi^{xx}$ is also approaching the Navier--Stokes solution, $\pi^{xx} = \frac{2\bar\eta\,s}{3\,\hat{x}^0(\tau,\eta)}$, as indicated by the dashed curves. When $\tau$ is fixed, the deviations from ideal limit are smaller at larger $|\eta|$. Nevertheless, this does not imply faster relaxation at larger rapidity region. 
The larger rapidity region has been initialized with smaller deviation from the long-time limit. To check the speed of relaxation, we may take the ratio between the time evolving values and their corresponding initial value, i.e., $\pi^{xx}(\tau,\eta) / \pi^{xx}(\tau_i,\eta)$ and $\Delta T(\tau,\eta) / \Delta T(\tau_i,\eta)$. Such ratios are shown in the lower panels of Fig.~\ref{fig2}. We observe that at large enough proper-time, the ratios are smaller at mid rapidity, which indicates faster relaxation. This is consistent with the fact that temperature is higher at the mid rapidity region, which means a smaller relaxation time.

\vspace{3mm}
\emph{Summary and Outlook.}
In this work, we derive new analytical solutions to the Boltzmann equation, which take the relaxation time approximation for the collisions kernel, associated with a newly discovered exact solution of ideal hydrodynamics~\cite{Shi:2022iyb}. The solution assumes homogeneity in the transverse plane, and allows non-trivial rapidity dependence. 
This is the first analytical solution of the Boltzmann equation that breaks the boost invariance, to the best of our knowledge.
With the distribution function, we further construct the stress tensor and compute the effective temperature and shear viscous stress tensor. We observe that both the temperature and the shear viscous stress tensor relax to the limit of hydrodynamics.
We also explicitly show the proper-time and rapidity dependence of the deviation of the temperature and shear viscous tensor from the long-time ideal limit, and we observe faster relaxation at mid rapidity, which is higher in temperature.
Our solution is useful in testing the applicability and accuracy of different approximations in the derivation of hydrodynamic equations. 
In particular, if taking the initial distribution in Eq.~\eqref{distributionf} to be anisotropic in momentum space, one may compare the results with those in anisotropic hydro~\cite{Florkowski:2010cf, Martinez:2010sc, Tinti:2013vba, Alqahtani:2017mhy}. Results will be reported in future publication.

\vspace{5mm}
\emph{Acknowledgement.} The authors thank Dr. Lipei Du for helpful discussion and Dr. Micheal Strickland for very valuable comments. This work is supported by Tsinghua University under grant Nos. 53330500923 and 100005024.

\begin{appendix}
\section{Long time asymptotic behavior}
Here provides appendix that analyzes the long-time asymptotic behavior~\eqref{eq:longtime} of the temperature and shear viscosity self-consistent equations~\eqref{eq:T_and_pi} in the main text. For convenience, we use $\tau$ to denote the hat proper time ($\hat x^0$) in this Appendix. For later convenience, we define that $F_{\bar\eta}(\tau) \equiv \frac{5\bar{\eta}}{e_{\bar\eta} T_i \tau_i}\big(\frac{\tau_i}{\tau}\big)^{\frac{2}{3}}$.
We also note that, when $\zeta \to 0$,
\begin{align}
\begin{split}
\tilde{\mathcal{H}}_0 (e^{-\zeta}) =\;& 
    \Big(1+\frac{8\zeta^2}{45}\Big)e^{-\frac{4}{3}\zeta} + \mathcal{O}(\zeta^3)\,,\\
\tilde{\mathcal{H}}_x (e^{-\zeta}) =\;& 
    \frac{8\zeta}{15\pi^2}e^{-\frac{11}{7}\zeta} + \mathcal{O}(\zeta^3)\,.
\end{split}
\end{align}

As the first attempt, we start from solving the ``zeroth iteration'' that putting the ideal solution on the right hand side of the integral, 
\begin{align}
\begin{split}
    \hat T_\mathrm{eff,0}^4(\tau)
=\;& 
    \frac{1}{5\bar{\eta}} \int_{\tau_i}^{\tau} D_\mathrm{ideal}(\tau, t) \tilde{\mathcal{H}}_0 (t/\tau) \hat T_\mathrm{ideal}^5(t) \,dt\,,
\end{split}
\end{align}
which leads to
\begin{align}
    \hat T_\mathrm{eff}(\tau) = e_{\bar\eta}\, T_i \Big(\frac{\tau_i}{\tau}\Big)^{\frac{1}{3}} \big(1 + \phi\, F_{\bar\eta}(\tau)\big)\,,
\end{align}
with $e_{\bar\eta} = 1$ and $\phi = -\frac{1}{6}$. Such coefficients do not agree with the numerical solution of solving the complete self-consistent equation, but the power of the time dependence is correct. We, therefore, solve the self-consistent equation for the coefficients.
The decay kernel reads
\begin{align}
\begin{split}
D(\tau_2, \tau_1) =\;&
    \exp\Big[-\frac{3}{2\, F_{\bar\eta}(\tau_2)} \Big( 1 - \Big(\frac{\tau_1}{\tau_2}\Big)^{\frac{2}{3}}\Big)\Big]
    \Big(\frac{\tau_1}{\tau_2}\Big)^{\phi}\,,
\end{split}
\end{align}
and
\begin{align}
\begin{split}
    & D(\tau, \tau\,e^{-\zeta}) 
=
    \exp\Big[-\frac{3}{2\, F_{\bar\eta}(\tau)} \Big( 1 - e^{-\frac{2}{3}\zeta}\Big)\Big]
    e^{-\phi\,\zeta}
\\
    =\;& \Big(1 + \frac{\zeta^2}{3\,F_{\bar\eta}(\tau)}
    -\frac{2\,\zeta^3}{27\,F_{\bar\eta}(\tau)}
    +\frac{\zeta^4}{18\,F_{\bar\eta}^2(\tau)}
    \Big) 
    e^{-(\frac{1}{F_{\bar\eta}(\tau)}+\phi)\,\zeta} + \mathcal{O}(\zeta^4)
    \,.
\end{split}
\end{align}

We replace the integration variable $t = \tau\, e^{-\zeta}$ in the self-consistent equation~\eqref{eq:T_and_pi}, and it becomes
\begin{align}
\begin{split}
    &\big(1 + \phi\, F_{\bar\eta}(\tau)\big)^4
\\=\;& 
    \frac{1}{F_{\bar\eta}}\int_{0}^{\infty} d\zeta\,
    \Big(1 + \frac{\zeta^2}{3\,F_{\bar\eta}(\tau)}
    -\frac{2\,\zeta^3}{27\,F_{\bar\eta}(\tau)}
    +\frac{\zeta^4}{18\,F_{\bar\eta}^2(\tau)}\Big) 
\\&\qquad\times
    e^{-\big(\frac{1}{F_{\bar\eta}}+\frac{2}{3}+\phi\big)\zeta}
    \Big(1+\frac{8\zeta^2}{45}\Big) 
    \big(1 + \phi\, F_{\bar\eta}(\tau)\big)^5
\\=\;& 
    1 + 4\phi\,F_{\bar\eta}(\tau)
    + \Big(\frac{16}{45} + \frac{8\phi}{3} + 6\phi^2 \Big) F_{\bar\eta}^2(\tau) + \mathcal{O}(F_{\bar\eta}^3)\,.
\end{split}
\end{align}
Thus, we find $\phi = -\frac{2}{15}$, and
\begin{align}
    \hat T_\mathrm{eff}(\tau) = e_{\bar\eta}\, T_i \Big(\frac{\tau_i}{\tau}\Big)^{\frac{1}{3}} - \frac{2\,\bar\eta}{3} \frac{\tau_i}{\tau} + \mathcal{O}(\tau^{-\frac{5}{3}})\,,
\end{align}

Then, computing the long time behavior of the shear stress tensor become straightforward,
\begin{align}
\begin{split}
    \hat \pi^{xx}(\tau)
=\;& 
    \frac{1}{5\bar{\eta}} \int_{\tau_i}^{\tau} d{t} D(\tau, t) \tilde{\mathcal{H}}_x (t/\tau) \hat T_\mathrm{eff}^5(t)
\\=\;& 
    \frac{8 e_{\bar\eta}^4 T_i^4}{15\pi^2}
    \Big(\frac{\tau_i}{\tau}\Big)^{\frac{4}{3}}
    F_{\bar\eta}(\tau)
    \Big(1- \frac{22\,F_{\bar\eta}(\tau)}{105}\Big)
    + \mathcal{O}(\tau^{-2})\,,
\end{split}
\end{align}
so that
\begin{align}
\begin{split}
    \frac{\hat \pi^{xx}(\tau)\,\tau}{\hat s(\tau)} =  \frac{2\,\bar\eta}{3} + \frac{40}{63}\frac{\bar{\eta}^2}{e_{\bar\eta} T_i \tau_i}\big(\frac{\tau_i}{\tau}\big)^{\frac{2}{3}}  + \mathcal{O}(\tau^{-\frac{4}{3}})\,.
\end{split}
\end{align}
As shown in the main text, such long time asymptotic formulae are in good agreement with the numerical solution of the self consistent equations~\eqref{eq:T_and_pi}.
\end{appendix}

\bibliographystyle{apsrev4-2}
\bibliography{ref}

\end{document}